\documentclass[12pt, a4paper, oneside]{article}

\usepackage[latin1]{inputenc}

\usepackage{latexsym,amsmath,amsthm,amsfonts,amssymb}
\usepackage{enumerate,array}
\usepackage{longtable,verbatim,wrapfig}
\usepackage{booktabs,colortbl,setspace}
\usepackage{cancel,anysize}
\usepackage{graphicx}
\usepackage{here}
\usepackage{url}
\usepackage{ragged2e}
\usepackage{multicol}

\usepackage[left=2.5cm,top=2cm,right=2.5cm,bottom=2.5cm]{geometry} 
\usepackage{geometry} 
\usepackage{anysize}
\usepackage{epsfig,graphicx,enumitem}
\usepackage{float}
\usepackage[footnotesize]{caption}
\usepackage[absolute]{textpos}

\newtheorem{teo}{Definition}
\usepackage{authblk}
\title{Adaptative significance levels in normal mean hypothesis testing}
\author[1]{Alejandra Estefanía Patiño Hoyos}

\author[2]{Victor Fossaluza}

\affil[1]{Institute of Mathematics and Statistics, University of São Paulo, Brazil; alejaeph@ime.usp.br}
\affil[2]{Institute of Mathematics and Statistics, University of São Paulo, Brazil; victorf@ime.usp.br}

\date{}                     
\begin{document}

\maketitle

\begin{abstract}

	The Full Bayesian Significance Test (FBST) for precise hypotheses was presented by Pereira and Stern (1999) as a Bayesian alternative instead of the traditional significance test based on \textit{p-value}. The FBST uses the evidence in favor of the null hypothesis ($H_0$) calculated as the complement of the posterior probability of the highest posterior density region, which is tangent to the set defined by $H_0$. An important practical issue for the implementation of the FBST is the determination of how large the evidence must be in order to decide for its rejection. In the Classical significance tests, the most used measure for rejecting a hypothesis is \textit{p-value}. It is known that \textit{p-value} decreases as sample size increases, so by setting a single significance level, it usually leads $H_0$ rejection. In the FBST procedure, the evidence in favor of $H_0$ exhibits the same behavior as the \textit{p-value} when the sample size increases. This suggests that the cut-off point to define the rejection of $H_0$ in the FBST should be a sample size function. In this work, we focus on the case of two-sided normal mean hypothesis testing and present a method to find a cut-off value for the evidence in the FBST by minimizing the linear combination of the type I error probability and the expected type II error probability for a given sample size.
\end{abstract}

\section{Introduction}

Oliveira (2014), motivated by Pereira (1985) suggests that the level of significance in hypothesis testing should be a sample size ($n$) function, in order to solve the problem of testing hypotheses in the usual way, in which change the $n$ influences the  null hypothesis probability rejection or acceptance. Instead of setting a single level of significance, Oliveira (2014) proposes to fix the type I and type II error probabilities weight ratio based on the incurred losses in each case, and thus, given a sample size, to define the level of significance that minimizes the linear combination of the decision errors probabilities. Oliveira (2014) showed that with this procedure, increasing the sample size implies that the probabilities of both kind of errors and their linear combination decrease, when in most cases, setting a single level of significance, independent of sample size, only type II error probability decreases. In the tests proposed by Oliveira (2014) the same conceptual basis of the usual tests for simple hypotheses is used, starting from Neyman-Pearson Lemma to find optimal and more powerful tests in which the error probabilities is minimized (DeGroot, 1975). This idea is extended to composite and sharp hypotheses, according to Pereira's initial work. In the sharp hypotheses cases, the FBST (Pereira and Stern, 1999) was implemented. 
\vspace{0.2cm}

Following the concepts in DeGroot (1975) and ${\textrm{Pereira (1985)}}$ associated to optimal hypothesis testing, as well as the conclusions in Oliveira (2014) regarding the relation between the level of significance and the sample size, we present a method to find a cut-off value $k$ for the evidence in the FBST as a function of $n$, this is $k=k(n)$ with $k \in [0,1]$, by minimizing the linear combination of the type I error probability and the expected type II error probability $a\alpha_{\varphi}+b\bar{\beta}_{\varphi}$. We will focus on the case of two-sided normal mean hypothesis testing.

\section{Methodology}

\begin{teo}

Let $f(\theta\vert x)$ be the posterior density of $\theta$ given the observed sample. Consider a sharp hypothesis ${H_0: \theta \in \Theta_0}$ and let ${T_{x}=\left\lbrace \theta \in \Theta: f(\theta\vert x)>sup_{\Theta_{0}}f(\theta\vert x) \right\rbrace}$ be the set tangential to $\Theta_0$. The measure of evidence in favor $H_0$ is defined as ${ev\left(\Theta_{0};x\right)=1-P(\theta \in T_{x}\vert x)}$. The \textbf{FBST} is the procedure that rejects $H_0$ whenever $ev\left(\Theta_{0};x\right)$ is small (Pereira et al., 2008).

\end{teo}

Suppose that $X_1,...,X_n$ are $c.i.i.d$  $Normal(\theta,\sigma^2)$ given $\theta$ ($\theta \in \mathbb{R}$ and $\sigma^2>0$ known), and define $X=(X_1,...,X_n)$. Let $\bar{X}=\sum_{i=1}^{n}X_i/n$ be a sufficient statistic for $\theta$, then, $\bar{X}\vert\theta \sim Normal(\theta,\sigma^2/n)$. Suppose also that $\theta \sim Normal(m,v^2)$ ($m \in \mathbb{R}$ and $v^2>0$). Then, the posterior distribution of $\theta$ given that $\bar{X}=\bar{x}$, is a normal distribution with parameters $(\sigma^2 m+nv^2\bar{x})/(\sigma^2+nv^2)$ and $(\sigma^2v^2)/(\sigma^2+nv^2)$. \\

Suppose that we wish to test the hypotheses

\begin{eqnarray*}
H_{0}&:& \theta=\theta_{0}\\
H_{1}&:& \theta\neq\theta_{0}
\end{eqnarray*}

Then, with $\Theta_0=\lbrace\theta_0\rbrace$, $T_{x}=\left\lbrace \theta \in \Theta: f(\theta\vert x)>f(\theta_{0}\vert x) \right\rbrace$.

\vspace{15pt}

Consider $\varphi(x)$ as the test such that

\begin{equation*}
\varphi(x)= \left\{ \begin{array}{l}  0, \quad if \quad ev\left(\Theta_{0};x\right)> k\\ \\ 1, \quad if \quad ev\left(\Theta_{0};x\right)\leq 
k. \end{array} \right.\;
\end{equation*}

\newpage
\begin{itemize}
\item Evidence 
\end{itemize}

\begin{figure}[H]
\setlength{\tabcolsep}{3pt}
\begin{tabular}{cc}
  \includegraphics[scale=0.6]{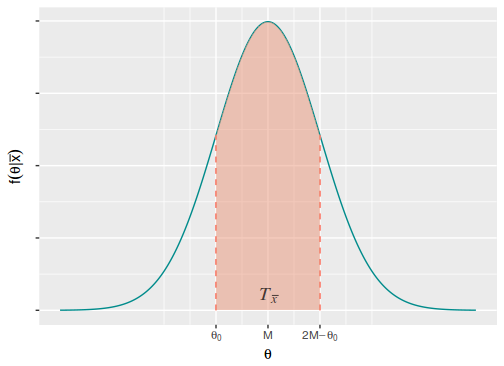} & \includegraphics[scale=0.6]{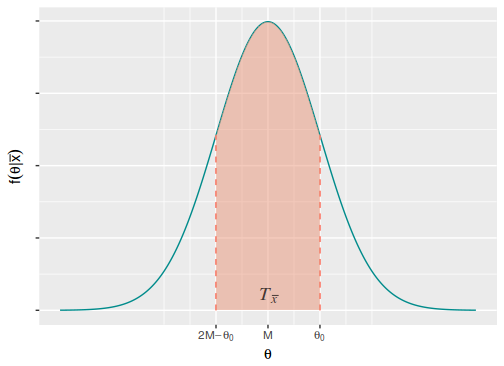} \\
 (a) \small $M>\theta_0$ & (b) \small  $M<\theta_0$
\end{tabular}
\caption{\small Tangential set to $\Theta_0$. $M=(\sigma^2 m+nv^2\bar{x})/(\sigma^2+nv^2)$.}
\label{evNN}
\end{figure}

\vspace{-10pt}
\begin{small}
\begin{eqnarray*}
ev\left(\Theta_{0};\bar{x}\right)&=&1-P\left( \theta \in T_{\bar{x}} \, \vert \bar{x}\right) \\
&=&1-P\left( \theta_0 \leq \theta \leq 2\left(\left.\dfrac{\sigma^2 m+nv^2\bar{x}}{\sigma^2+nv^2}\right) -\theta_0 \, \right\vert \bar{x}\right)\\ \\
&=& 2 \, \Phi\left( -\dfrac{\Bigl| \sigma^2(\theta_0-m)+nv^2(\theta_0-\bar{x})\Bigr|}{\sigma v\sqrt{\sigma^2+nv^2}}\right),
\end{eqnarray*}
\end{small}

where $\Phi$ is the standard normal cumulative distribution function.


\begin{itemize}
\item Power function
\end{itemize}
\begin{small}
\begin{flalign*}
\pi_{\varphi}(\theta) &=P(ev\left(\Theta_{0};\bar{X}\right) \leq k \vert \theta)&\\
&= P\left( 2 \, \Phi\left( -\dfrac{\Bigl| \sigma^2(\theta_0-m)+nv^2(\theta_0-\bar{x})\Bigr|}{\sigma v \sqrt{\sigma^2+nv^2}}\right)\leq k \Biggm| \theta\right)&\\
&=1-P\left(\Biggl| \dfrac{\sigma(\theta_0-m)}{\sqrt{n}\,v^2}-Z-\dfrac{(\theta-\theta_0)}{\sigma/\sqrt{n}}\Biggr| \leq \left. -\underbrace{\dfrac{\sqrt{\sigma^2+nv^2}\, \Phi^{-1}\left( \dfrac{k}{2}\right)}{\sqrt{n}\,v}}_{Q}  \right \vert \theta\right)& \\
&= 1-P\left(Q \leq -\dfrac{\sigma(\theta_0-m)}{\sqrt{n}\,v^2}+Z+\dfrac{(\theta-\theta_0)}{\sigma/\sqrt{n}} \leq -Q  \Biggm| \theta\right)&\\
&= 1-P\left(\underbrace{Q+\dfrac{\sigma(\theta_0-m)}{\sqrt{n}\,v^2}-\dfrac{(\theta-\theta_0)}{\sigma/\sqrt{n}}}_{z_1} \leq Z \leq \underbrace{-Q+\dfrac{\sigma(\theta_0-m)}{\sqrt{n}\,v^2}-\dfrac{(\theta-\theta_0)}{\sigma/\sqrt{n}}}_{z_2}  \Biggm| \theta\right)&\\\\
&=1-\left[ \Phi(z_2)-\Phi(z_1)\right\vert\theta].&  
\end{flalign*}
\end{small}

\begin{itemize}
\item Type I error probability 
\end{itemize}
\begin{small}
\begin{flalign*}
\alpha_{\varphi} &=P(ev\left(\Theta_{0};\bar{X}\right)\leq k \vert \theta=\theta_0)&\\\\
&= 1-P\left(\underbrace{\dfrac{\sqrt{\sigma^2+nv^2}\, \Phi^{-1}\left( \dfrac{k}{2}\right)}{\sqrt{n}\,v}+\dfrac{\sigma(\theta_0-m)}{\sqrt{n}\,v^2}}_{z_{1}^{*}} \leq Z \leq \underbrace{-\dfrac{\sqrt{\sigma^2+nv^2}\, \Phi^{-1}\left( \dfrac{k}{2}\right)}{\sqrt{n}\,v}+\dfrac{\sigma(\theta_0-m)}{\sqrt{n}\,v^2}}_{z_{2}^{*}}  \right)&\\\\
&= 1-\left[ \Phi(z_{2}^{*})-\Phi(z_{1}^{*})\right].&
\end{flalign*}
\end{small}

\begin{itemize}
\item Expected Type II error probability 
\end{itemize}
\begin{small}
\begin{eqnarray*}
\bar{\beta}_{\varphi} &=& E\left[ 1-\pi(\theta)\vert \theta \in \Theta_{1}\right]  \\
&=& E\left[ \Phi(z_{2})-\Phi(z_{1})\vert \theta \in \Theta_{1}\right]\\
&=&\int_{\Theta \setminus\lbrace\theta_{0}\rbrace}\left[ \Phi(z_{2})-\Phi(z_{1})\right] f(\theta)d\theta.\\ 
\end{eqnarray*}
\end{small}

\section{Results}

\begin{figure}[H]
\setlength{\tabcolsep}{-6pt}
\begin{tabular}{cc}
    \includegraphics[scale=0.7]{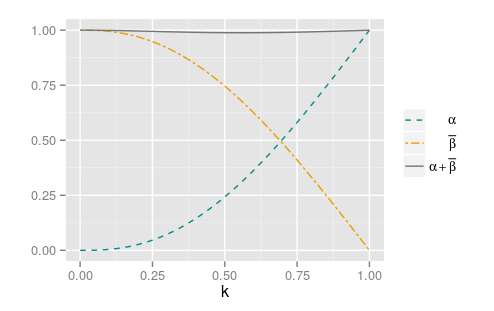} &\includegraphics[scale=0.7]{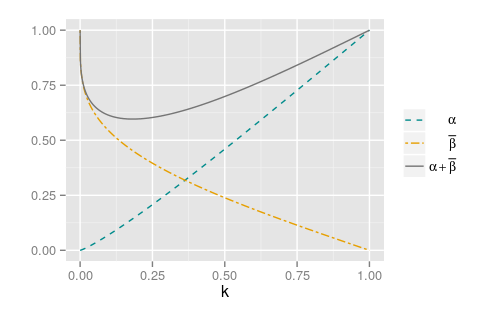} \\
\small  (a) \, $\theta \sim Normal(0,0.1)$  & \small (b) \,   $ \theta \sim Normal(0,1)$ \\[0pt]
 \end{tabular}
\caption{Error probabilities ($\alpha$, $\bar{\beta}$ and $\alpha+\bar{\beta}$) as functions of $k$. $H_{0}:\theta=0$ vs. $H_{0}:\theta\neq0$, $n=50$,  $a=b=1$.}
\end{figure}

\setlength\columnsep{-20pt}
\begin{multicols}{2}
\begin{table}[H]
\centering
\begin{small}
\begin{tabular}{crc}
  \hline\hline
  &$k$  & \\ 
\cline{2-3} 
 $n$  & $v^2=0.1$ & $v^2=1$  \\ 
  \hline 
  10 & 0.76244 & 0.40574  \\ 
  50 & 0.55893 & 0.18178  \\ 
  100 & 0.46904 & 0.12234 \\ 
  150 & 0.42266 & 0.09651  \\ 
  200 & 0.39316 & 0.08142 \\
  250 & 0.37209 & 0.07131 \\ 
  300 & 0.35591 & 0.06398  \\ 
  350 & 0.34290 & 0.05838  \\ 
  400 & 0.33198 & 0.05395  \\ 
  450 & 0.32256 & 0.05033  \\ 
  500 & 0.31433 & 0.04732  \\ 
  1000 & 0.26226 & 0.03174  \\
  1500 & 0.23230 & 0.02533 \\
  2000 & 0.21119 & 0.02168 \\ 
   \hline\hline
\end{tabular}
\end{small}
\caption{\scriptsize Cut-off values $k$ for $ev\left(0;\bar{x}\right)$ as function of $n$, with $\theta \sim Normal(0,0.1)$ and $\theta \sim Normal(0,1)$. $H_{0}:\theta=0$ vs. $H_{0}:\theta \neq 0$, $a=b=1$.}
\label{tabknpriorisNN}
\end{table}
\columnbreak
\begin{figure}[H]
\centering
\includegraphics[scale=0.6]{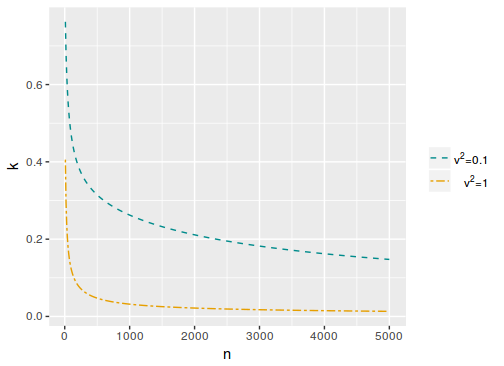}
\caption{ \scriptsize Cut-off values $k$ for $ev\left(0;\bar{x}\right)$ as function of $n$, with $\theta \sim Normal(0,0.1)$ and $\theta \sim Normal(0,1)$. $H_{0}:\theta=0$ vs. $H_{0}:\theta \neq 0$, $a=b=1$.}
\label{knpriorisNN}
\end{figure}
\end{multicols}

By increasing $n$, $k$ shows a decreasing trend, which leads us to interpret that the influence of sample size on the determination of the cut-off for $ev\left(0;\bar{x}\right)$ is very relevant.\\

Also, it is possible to see the differences between the two results with the different kind of prior distributions, and therefore, to identify their importance at the moment of defining the cut-off value for $ev\left(0;\bar{x}\right)$. It can also be observed that, when the prior is  less informative, the $k$ value is smaller.

\begin{figure}[H]
\setlength{\tabcolsep}{-6pt}
\begin{tabular}{cc}
    \includegraphics[scale=0.7]{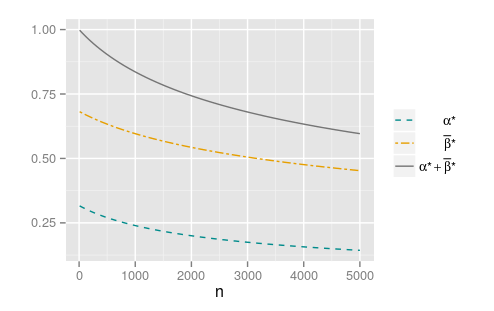} &\includegraphics[scale=0.7]{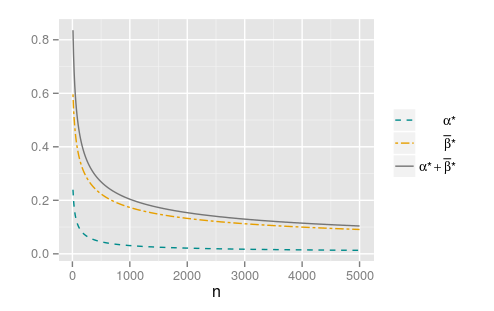} \\
\small  (a) $ Normal(0,0.1)$  & \small (b)   $ Normal(0,1)$ \\[0pt]
\end{tabular}
 \begin{sloppypar} \caption{Error probabilities for the optimal $k$ ($\alpha^{*}$, $\bar{\beta^{*}}$ and $\alpha^{*}+\bar{\beta^{*}}$) as function of $n$.  ${H_{0}:\theta=0}$ vs. $H_{0}:\theta\neq0$, $a=b=1$. }  
\end{sloppypar}
\label{kn11}
\end{figure}

As the sample size increases, the probabilities of both types of errors and their linear combination decrease.

\section*{References}

\begin{description}

{\item[] DeGroot, M. H. (1975). 
Probability and Statistics, 2nd edn, {\it Addison-Whesley Publishing Company, Massachusetts.}}

{\item[] Oliveira, M. C. (2014). 
{\it Definição do nível de significância em função do tamanho amostral}. Dissertação de Mestrado, Universidade de São Paulo, Instituto de Matemática e Estatística. Departamento de Estatística, São Paulo.}

{\item[] Pereira, C. A. B., Stern, J. M. and Wechsler, S. (2008). 
an a significance test be genuinely bayesian?. {\it Bayesian Analysis} {\bf 3} (1), 79-100.}

{\item[] Pereira, C. A. B. (1985). 
{\it Teste de hipóteses definidas em espaços de diferentes dimensões: visão Bayesisana e \\interpretação Clássica}. Tese de Livre Docência, Universidade de São Paulo, Instituto de Matemática e Estatística. Departamento de Estatística, São Paulo.}

{\item[] Pereira, C. A. B. and Stern, J. M. (1999). 
Evidence and credibility: Full bayesian significance test for precise hypotheses. {\it Entropy} {\bf 1} (4), 99-110.}

\end{description}

\end{document}